\documentclass{optica-article}

\journal{opticajournal} 

\articletype{Research Article}

\usepackage{lineno}

\usepackage{xcolor}
\usepackage{soul}

\usepackage{comment}

\begin{document}

\title{Noise resilient real-time phase imaging via undetected light}

\author{\author{Josué R. León-Torres,\authormark{1, 2, 3*} Patrick Hendra,\authormark{1, 2} Yugant Mukeshbhai Hadiyal,\authormark{1} Christopher Spiess,\authormark{2} Fabian Steinlechner,\authormark{1, 2} Frank Setzpfandt,\authormark{1, 2} Markus Gräfe,\authormark {2, 4} and Valerio Flavio Gili\authormark{2}}}

\address{
\authormark{1}Abbe Center of Photonics, Friedrich Schiller University Jena, Albert-Einstein-Straße 6, 07745 Jena, Germany\\

\authormark{2}Fraunhofer Institute for Applied Optics and Precision Engineering IOF, Albert-Einstein-Straße 7, 07745 Jena, Germany\\

\authormark{3}Cluster of Excellence Balance of the Microverse, Friedrich Schiller University Jena, Jena, Germany\\

\authormark{4}Institute for Applied Physics, Technical University of Darmstadt, Otto-Berndt-Straße 3, 64287 Darmstadt, Germany
}

\email{\authormark{*}josue.ricardo.leon.torres@iof.fraunhofer.de} 

\begin{abstract*} 
Quantum imaging with undetected light has recently emerged as a technique in which quantum correlations and nonlinear interferometry are combined to decouple illumination and detection paths. This approach has been more recently extended and combined with digital phase-shifting holography and off-axis holography to extract both the amplitude and phase information of a sample relying on single-photon interference. Despite these advantages, implementing the technique in real-world scenarios where the observed system is subject to environmental noise and dynamic variations remains challenging. The primary limitation lies in the inability of quantum imaging systems to retrieve object information in real time under high-noise conditions. Here, we experimentally demonstrate real-time amplitude and phase imaging in noisy environments, building upon our previous implementation of quantum off-axis holography. Our results demonstrate real-time imaging at acquisition rates up to 4~Hz, even when the noise level exceeds the signal by an order of magnitude.

\end{abstract*}

\section{Introduction}

Quantum imaging has demonstrated its potential across a variety of scenarios\cite{Jorge_Review_2024, Defienne2024}, including microscopy by surpassing classical limitations \cite{tenne_super-resolution_2019, ortolano_quantum_2023, Silberhorn}, imaging under strong environmental background by exhibiting remarkable resilience to noise \cite{gregory_imaging_2020, gregory_noise_2021, defienne_quantum_2019, fuenzalida2023, szuniewicz_noise-resistant_2023}, and sensing at sub-shot-noise levels by exploiting photon-pair correlations \cite{Genovese, ruo-berchera_improving_2020, samantaray_realization_2017, brida_experimental_2010}. Recent developments such as quantum illumination (QI) \cite{defienne_quantum_2019, gregory_imaging_2020} and quantum imaging with undetected light (QIUL) \cite{Marta1, Sebastian2022, fuenzalida2023} have further introduced noise-resilient schemes capable of retrieving both amplitude and phase information under challenging conditions. However, many of these protocols still require long integration times, coincidence detection, or extensive statistical post-processing, rendering them unsuitable for investigating dynamic samples that evolve on short time scales.

A crucial factor for integrating quantum imaging and sensing technologies into real-world applications is the ability to perform real-time amplitude and phase imaging of dynamic samples in realistic deployment conditions, where systems are exposed to environmental background noise. Achieving this capability would open new opportunities across diverse fields, including biomedical imaging in the life sciences, quality control in industrial production lines, vision systems for autonomous vehicles (e.g., LiDAR), and security and surveillance. However, current techniques lack this real-time capability. Existing quantum imaging approaches designed to operate in noisy environments typically require exposure times ranging from several minutes to hours for coincidence-based detection protocols \cite{gregory_imaging_2020, gregory_noise_2021, defienne_quantum_2019, szuniewicz_noise-resistant_2023}, and on the order of tens of seconds for single-photon detection schemes \cite{fuenzalida2023}. To the best of our knowledge, no existing method enables real-time amplitude and phase imaging under noisy environmental conditions.

Among these approaches, QIUL stands out by enabling image formation through undetected photons. QIUL is an interferometric imaging technique in which one photon of a correlated pair probes the object, while its partner, which has never interacted with the sample is detected and used to reconstruct the image \cite{Lemos2014, Inna2020, Kutas2022, Marta1, Sebastian2022, Topfer:25, Leon-Torres(2024), Pearce2024, Kumar2025}. This process allows simultaneous retrieval of both amplitude and phase information without direct detection of the photon that interacted with the object.

Building on the recently introduced quantum imaging technique that combines QIUL and Fourier off-axis holography (OAH) \cite{Leon-Torres(2024)}, we investigate its feasibility for real-time phase reconstruction under dynamic and noisy conditions. This approach, environmental noise is emulated by superimposing a classical light source onto the quantum image at the detection plane.
Our implementation enables the retrieval of both amplitude and phase information from a single-shot measurement \cite{Leon-Torres(2024), Topfer:25, Pearce2024}, thereby drastically reducing the acquisition time. Fourier off-axis holography provides the foundation for this noise-resistant imaging approach. We introduce a continuous-wave laser as a controllable classical noise source to assess the robustness of the technique, and incorporate a spatial light modulator (SLM) in the undetected-photon path to generate time-varying phase profiles. We demonstrate that the holographic phase reconstruction remains stable and accurate even under strong noise and rapidly changing phase profiles, thereby opening new opportunities for practical applications of QIUL where both speed and noise resilience are essential.

\section{Theory}

\subsection{Noise resilience mechanism}

The noise resilience of the approach is enabled by quantum off-axis holography with undetected light (QOAHUL) \cite{Leon-Torres(2024), Pearce2024, Topfer:25} an interferometric quantum imaging scheme that integrates Fourier OAH \cite{Sanchez, Kim, Verrier, Nguyen, Goodman-Book, Schnars2015}, as the foundation of our noise-resilient phase-retrieval technique. 

\begin{figure}[ht!]
\centering\includegraphics[width=8.5cm]{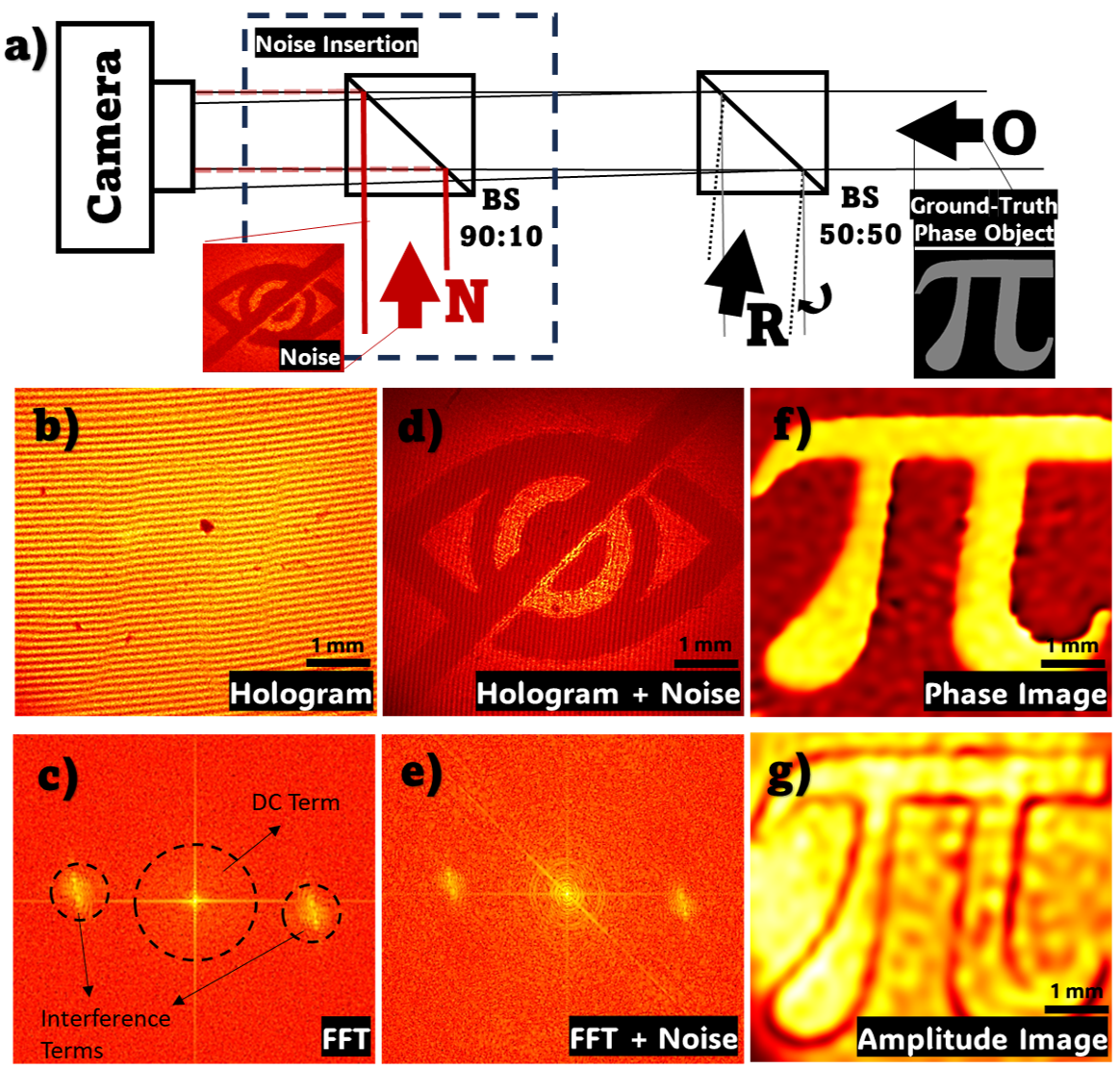}
\caption{Quantum image distillation.
(a) Simplified schematic of the imaging system based on Fourier OAH.
(b) Quantum hologram in the absence of noise.
(c) Fourier transform of the noise-free hologram.
(d) Hologram corrupted by the superposition of classical noise.
(e) Fourier transform of the noisy hologram.
(f) Reconstructed phase image from the noisy hologram.
(g) Corresponding reconstructed amplitude image.} 
\label{fig:Distillation}
\end{figure}

In Fourier OAH, a hologram is formed by the interference between an object field $O(x,y)$ and a tilted reference field $R(x,y)$, recorded as an intensity pattern at the detection plane. This recorded intensity encodes the complex amplitude of the object wave through spatially modulated interference fringes \cite{Goodman-Book, Schnars2015}. A simplified schematic of this configuration is shown in the right panel of Fig.~(\ref{fig:Distillation}a), with the corresponding ground-truth pattern ($\pi$ symbol) displayed as an inset.
To describe the formation of the hologram, we introduce the intensity distribution recorded at the camera plane \cite{Sanchez, Verrier, Cuche}:
\begin{equation}
I(x,y) =  \underbrace{|O(x,y)|^{2} + |R(x,y)|^{2})}_{\text{DC term }} 
        + \underbrace{O(x,y)R^*(x,y) + O^*(x,y)R(x,y)}_{\text{Interference terms }}.
\end{equation}

The first two terms in Eq.~(1) correspond to the squared amplitudes of the object and reference fields, representing their respective intensities. The last two terms arise from the interference between these fields and carry the amplitude and phase information of the object \cite{Sanchez, Kim, Nguyen}. Together, these terms form the hologram, as illustrated in Fig.~(\ref{fig:Distillation}b). The asterisk symbol (\textasteriskcentered) denotes the complex conjugate operation.
Examining the hologram in the Fourier domain reveals that the different terms of Eq.~(1) occupy distinct spatial-frequency regions, as shown in Fig.~(\ref{fig:Distillation}c).



We now consider the effect of an additional noise contribution on the hologram formation. Once a noise source is introduced into the imaging system, as depicted on the left side of Fig.~(\ref{fig:Distillation}a), the hologram quality degrades. The corresponding noise image, generated by a classical light source that emulates the environmental background, is shown as an inset. Equation~(2) describes the intensity spatial distribution of the hologram corrupted by the presence of noise in real space. The noise $N(x,y)$ hinders the image reconstruction process in real space by masking the object information. In the recorded hologram, the noise term appears solely as an additional intensity background, as illustrated in Fig.~(\ref{fig:Distillation}d). Because the noise lacks phase coherence with either the object or reference fields, it contributes only as an additive background intensity, thereby reducing the contrast of the interference fringes in the hologram:
\begin{equation}
I(x,y) = |O(x,y)|^2 + |R(x,y)|^2 +  \underbrace{|N(x,y)|^2}_{\text{Noise term}}
        + O(x,y)R^*(x,y) + O^*(x,y)R(x,y).
\end{equation}

A Fourier transform of Eq.~(2) reveals how the contributions from the beam intensities and interference terms are redistributed in Fourier space, as shown in Fig.~(\ref{fig:Distillation}e),
\begin{equation}
\begin{split}
\tilde{I}(k_x, k_y) = &
\underbrace{|\tilde{O}(k_x,k_y) |^2
          + |\tilde{R}(k_x,k_y) |^2 
          + |\tilde{N}(k_x,k_y) |^2}_{\text{DC term}} \\
&+ \tilde{O}(k_x, k_y) \otimes \delta(k_x - k_{rx}) \, \delta(k_y - k_{ry}) + \tilde{O}^*(k_x, k_y) \otimes \delta(k_x + k_{rx}) \, \delta(k_y + k_{ry}),
\end{split}
\end{equation}
where the tilde, $\tilde{f}(k_x,k_y)$ denotes the Fourier transform of $f(k_x,k_y)$, $\otimes$ represents convolution, $k_x$ and $k_y$ denote spatial frequencies in the horizontal and vertical axis, respectively. 
The last two terms of Eq.~(3) have a displacement proportional to the wavevector $\vec{k}=(k_{rx}, \ k_{ry})$, that is usually carried by the reference beam \cite{Sanchez, Verrier, Cuche}.
In the Fourier domain, the spatial-frequency components of the noise are spread over a broad range of frequencies with varying weights, rather than contributing uniformly as in the real-space intensity. As a result, the relative contribution of the noise spatial frequencies is reduced and the interference terms can be selectively filtered when the relative weights of the noise components are small compared to those associated with the object. Under these conditions, the phase and amplitude images encoded in the last terms of Eq.~(3) can be reconstructed, as shown in Figs.~(\ref{fig:Distillation}f) and~(\ref{fig:Distillation}g), respectively.

Our scheme exploits the interferometric modulation between the object and reference beams to encode the object information into the interference fringes. This modulation produces distinct spatial frequency components in Fourier space: the interference terms and the low-frequency background terms, which may include the DC and noise contributions \cite{Goodman-Book, Schnars2015}. When the noise level is small compared to the modulation imposed by the interference fringes, the object information can be reliably retrieved and the residual noise can be effectively suppressed through spectral filtering in Fourier space.

\section{Methods and Results}

\subsection{Experimental Setup}

\begin{figure}[ht!]
\centering\includegraphics[width=0.8\textwidth]{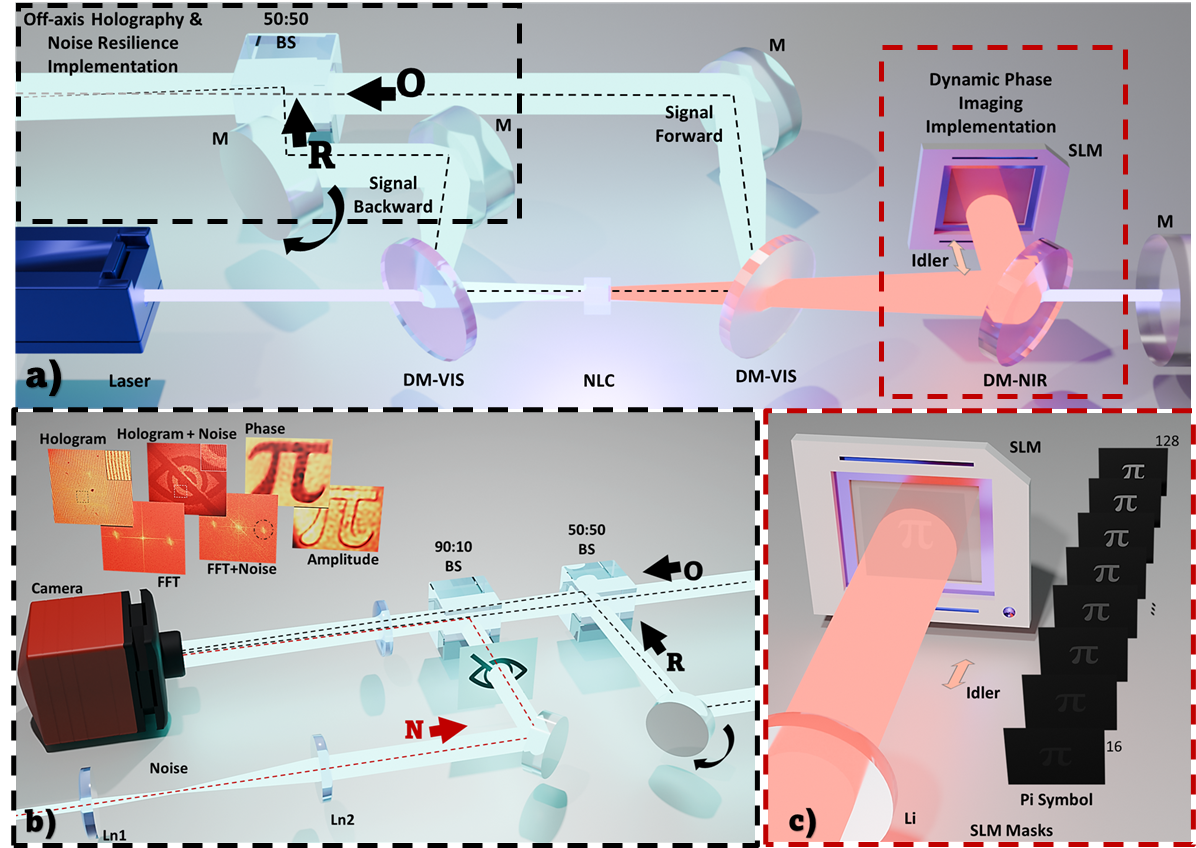}
\caption{Noise resilience real-time phase imaging implementation based on a hybrid-nonlinear interferometer. (a) Experimental setup: a CW laser at 405~nm pumps a 2-mm-long PPKTP crystal bidirectionally to generate photon pairs via SPDC, producing an idler beam at 730~nm (red) and a signal beam at 910~nm (cyan). (b) Quantum Off-axis holography with undetected light: the object and reference signal beams overlap at the camera plane. Fourier off-axis holography enables real-time image reconstruction despite the presence of noise. (c) Dynamic phase imaging implementation: the SLM projects a sequence of different phase patterns which are probed by the idler beam.}
\label{fig:Setup}
\end{figure}

The experimental setup consists of a hybrid-nonlinear interferometer \cite{Leon-Torres(2024), Kim2024}, in which a 2~mm long PPKTP crystal is pumped bidirectionally by a 405~nm continuous-wave (CW) laser. Signal (910~nm) and idler (730~nm) photons are generated via spontaneous parametric down conversion (SPDC) in both forward and backward directions. The forward- and backward- generated signal beams propagate through a Mach-Zehnder interferometer, whereas the corresponding idler beams propagate through a Michelson interferometer, see Fig.~(\ref{fig:Setup}a). The imaging system is arranged such that the Fourier plane of the crystal is projected onto the mirror in the pump path, onto the SLM in the idler path, and onto the 50:50 beam splitter (BS) in the signal paths. This configuration enables precise spatial control of the signal modes by introducing a small relative angle between them, thereby allowing the use of Fourier OAH as an image-processing method \cite{Leon-Torres(2024), Topfer:25, Pearce2024}; see Fig.~(\ref{fig:Setup}b). From this point onward, we refer to the forward-generated beams as the object beams and the backward-generated beams as the reference beams.

To emulate dynamic phase behavior, a Hamamatsu SLM was placed in the idler path, as shown in Fig.~(\ref{fig:Setup}c). The idler beam interacts with the SLM and is back-reflected to the crystal plane, thereby inducing coherence in its twin signal photon \cite{Mandel1991, Wiseman(2000), Lahiri2019}. This process enables the probability amplitudes of the forward- and backward-propagating signal photons to interfere, allowing the extraction of the object information. The SLM is programmed to project phase patterns with distinct features and phase values ranging from $0$ to $\pi$ at different update rates.

Additionally, a CW laser at 910~nm is employed as a classical noise source.Prior to entering the detection path, the noise beam is expanded and passed through a binary object, which is then imaged onto the camera plane. The noise beam (red dashed line) is subsequently combined with the object and reference signal beams via a 90:10 BS, resulting in spatial overlap at the camera plane. Despite never interacting with the SLM the signal beams are used for detection and enables retrieval of the phase information imprinted on the SLM. A schematic of this implementation is shown in Fig.~(\ref{fig:Setup}b).


Our experimental setup enables the extraction of both amplitude and phase information in a single-shot measurement. The integration of the noise source and SLM allows us to perform real-time phase imaging under noisy environmental conditions. This represents a significant improvement over previous implementations based on multi-frame acquisition \cite{fuenzalida2023, Sebastian2022, defienne_quantum_2019, gregory_imaging_2020, gregory_noise_2021}, for which real-time image reconstruction is not feasible.

\subsection{Noise resilient phase imaging}

To evaluate the resilience of the approach under noisy conditions, we vary the signal-to-noise intensity ratio by keeping the signal flux constant while progressively increasing the noise intensity until image reconstruction is no longer feasible. To this end, we first superimpose the classical noise beam and the quantum hologram at the camera plane. In this initial test, an unstructured noise beam is used; no object is placed in the noise path. 

The signal flux is characterized by the mean intensity value of the field of view (FoV) at the camera plane, considering only the signal photons generated in a single pass through the crystal. The forward and backward directions yield the same signal flux. An exposure time of 1~s is used for the measurement. Similarly, the mean noise intensity was measured under the same conditions.  The images acquired in this section have dimensions of 1024~×~1024~pixels. The SLM was configured to project the Greek letter $\pi$ with a gray value of 128, corresponding to a phase value of 3.14~rad.

\begin{figure}[ht!]
\centering\includegraphics[width=\textwidth]{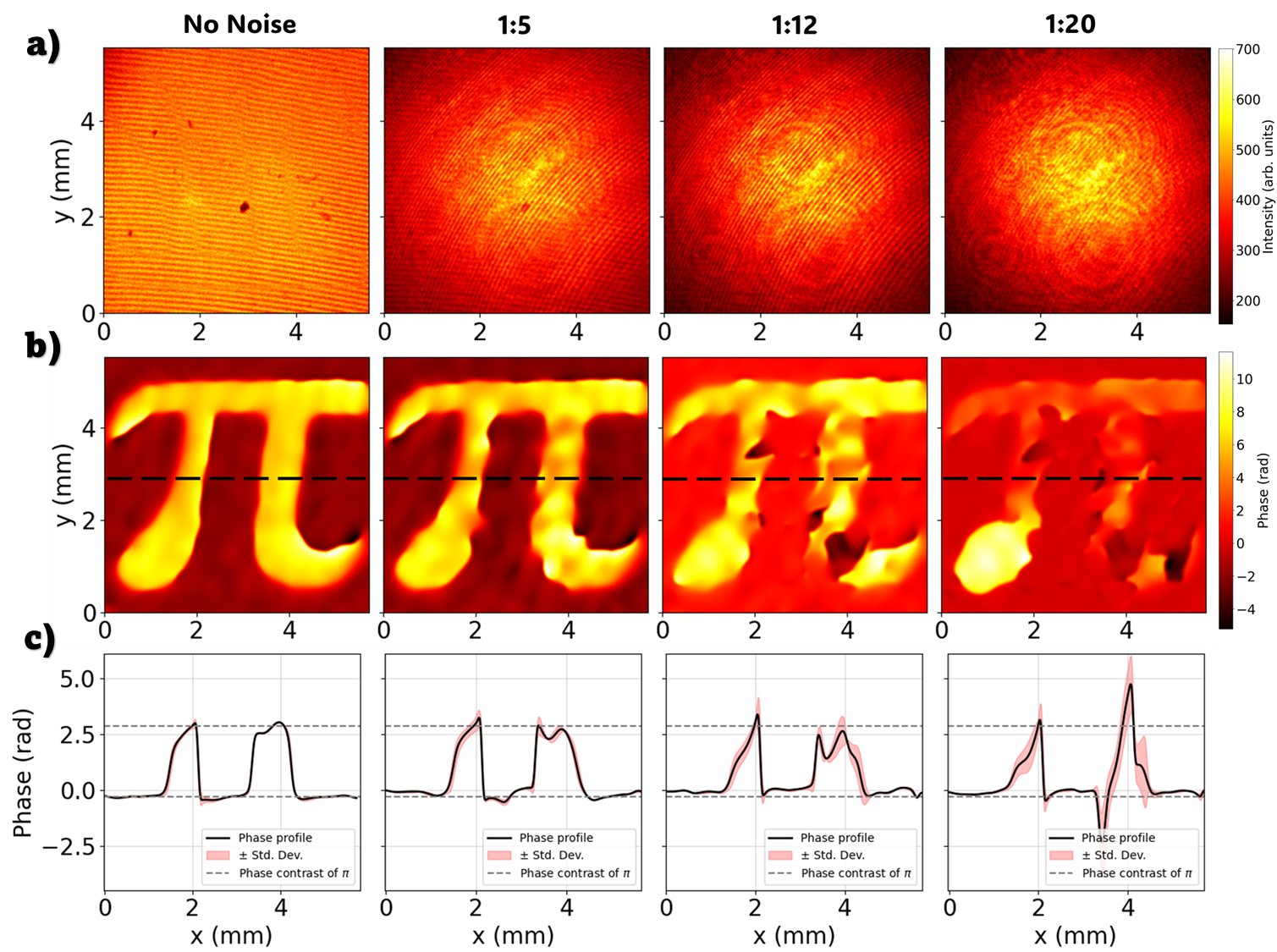}
\caption{Our noise-resilience method is evaluated by varying the ratio between the signal and noise intensities, ranging from the noise-free case to a ratio of 1:20, where the noise intensity is twenty times higher than the signal intensity. (a) Shows the overlap of the noise beam and the quantum hologram. (b) Reconstructed phase images of the Greek letter $\pi$, and (c) line profile plots extracted from the phase images (dashed black line) to evaluate the phase contrast at each signal-to-noise ratio.}
\label{fig:SNR}
\end{figure}

The three beams (object, reference, and noise) are superimposed at the camera plane, and the experimental results are presented in Fig.~(\ref{fig:SNR}a). The first row shows the superposition of the classical noise and quantum hologram images, where the noise intensity increases from left to right. The corresponding mean signal-to-noise intensity ratios are as follows: no noise, where the fringe pattern of the quantum hologram is clearly visible, and higher noise levels of 1:5, 1:12, and 1:20, where the presence of noise progressively obscures the hologram.

Figure ~(\ref{fig:SNR}b) shows the reconstructed phase images distilled from the noise using QOAHUL, corresponding to the superposed images in the first row. Phase reconstruction is successfully achieved in all cases. At higher noise levels, the sharpness of the object features decreases due to the overlap of the noise spatial frequencies with the interference terms in Fourier space. These effects are shown in Fig.~(\ref{fig:SNR}c). The plot profiles are obtained from a transverse cut of the phase images (dashed black line). The dashed gray horizontal lines indicate the theoretical phase contrast of 3.14~rad, while the red shaded region around the line profile represents the standard deviation of the phase values, calculated from the top and bottom 50 consecutive lines.

\begin{figure}[ht!]
\centering\includegraphics[width=0.5\textwidth]{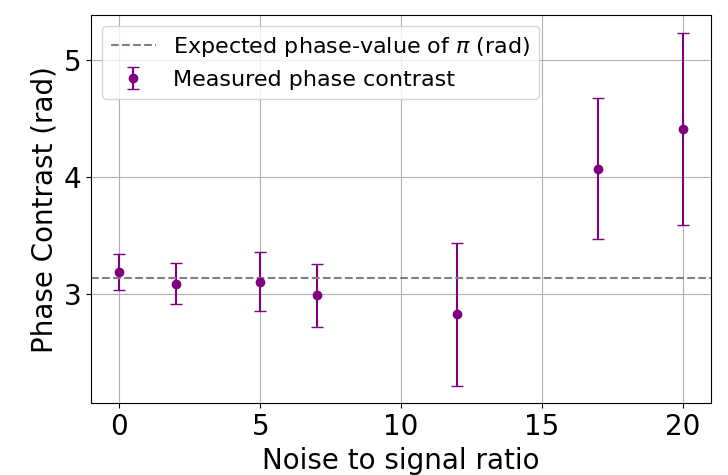}
\caption{Evolution of the phase contrast at different noise levels. The dashed-line represents the expected phase-value of $\pi$ rad.}
\label{fig:Quant1}
\end{figure}

Our measurements show that phase accuracy decreases as the noise intensity increases, as evidenced by the growing deviation and uncertainty of the phase contrast shown in Fig.~(\ref{fig:Quant1}). This behavior is consistent with the increased overlap of noise spatial frequencies with the interference terms in Fourier space and becomes particularly pronounced once the noise reaches seventeen and twenty times the signal intensity, where noticeable deviations from the expected phase value are observed. Although the phase contrast exhibits large uncertainties at these high noise levels, the original object features remain clearly recognizable, as shown in Fig.~(\ref{fig:SNR}b).
The phase-contrast values shown in Fig.~(\ref{fig:Quant1}) are extracted from three evenly spaced line profiles across each reconstructed phase image. From each profile, four phase values within the object region are sampled. The mean phase contrast and its standard deviation are then computed from these profiles.

\subsection{Spatial frequencies of the noise source}

Given the nature of our imaging scheme, which is based on Fourier off-axis holography, the spatial-frequency distribution of the noise source plays a critical role in the performance of our noise-resilience method. When the spectral distributions of the noise and the object strongly overlap, image reconstruction becomes unfeasible. However, our experimental implementation allows control over the fringe density and orientation, and consequently over the position of the spectral distribution of the object in Fourier space. This control is achieved by tuning the angle between the object and reference signal beams (see Fig.~(\ref{fig:Setup}b)). As long as the spectral distribution of the noise remains localized within a limited region of Fourier space, the fringe orientation can be adjusted so that the spectral components of the interference terms do not overlap with those of the noise.

\begin{figure}[ht!]
\centering\includegraphics[width=0.75\textwidth]{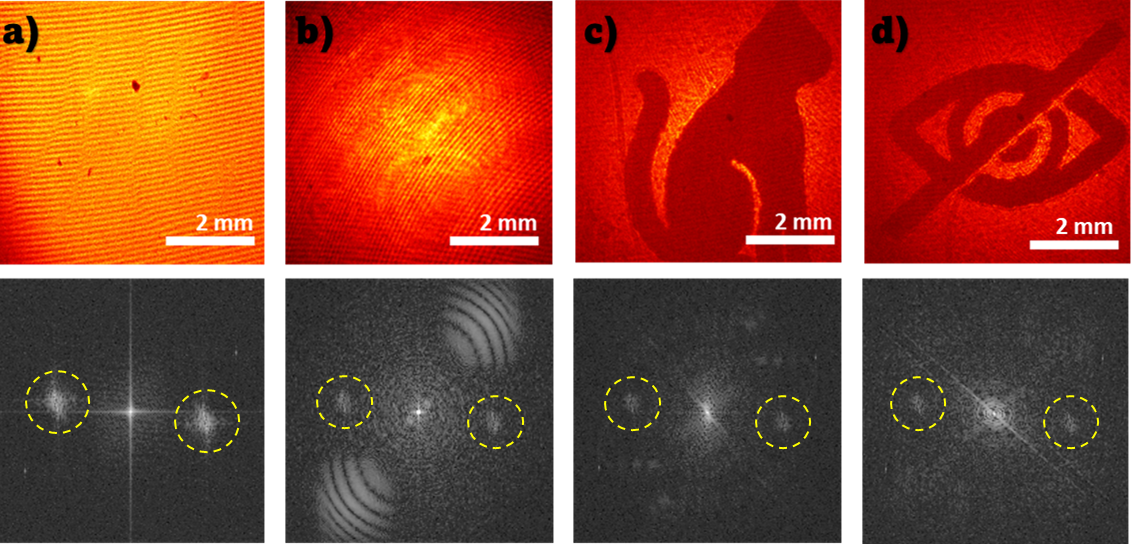}
\caption{ Overlap of the quantum hologram and noise beam profiles. (a) Overlap at the camera plane.(b) Corresponding overlap at the Fourier plane. In Fourier space, the interference terms can be positioned such that the overlap with the noise spatial frequencies is minimized, allowing smoother spectral filtering and improved image reconstruction.}
\label{fig:FFT}
\end{figure}

Four binary objects were used to demonstrate the effect of different spectral distributions of the noise source. Figure~(\ref{fig:FFT}) illustrates the capability of our scheme to retrieve the object information when the noise exhibits different Fourier components and the degree of overlap with the interference terms (enclosed in the yellow dashed-circles) varies. 

\subsection{Real-time phase imaging}

To demonstrate the noise resilience of our technique for real-time phase imaging, the SLM projects two mask sequences at rates of 1~Hz and 4~Hz, while the camera exposure time was fixed at 200~ms for both cases. Amplitude and phase images are reconstructed in real time, with and without noise. A signal-to-noise ratio of 1:5 was used in all measurements presented in this section (see the supplementary videos attached to this publication). 


To speed up the image acquisition process, the FFT image size is reduced from 1024~×~1024~pixels to 64~×~64~pixels for all measurements in this section. This reduction is performed during the FFT reconstruction step by cropping and re-centering the region containing the object-related spatial frequencies, which in all cases are confined within a 64~×~64~pixel area. This procedure preserves the full spatial resolution and phase-contrast information of the object.

\subsubsection{Dynamic features over time}

\begin{figure}[ht!]
\centering\includegraphics[width=0.8\textwidth]{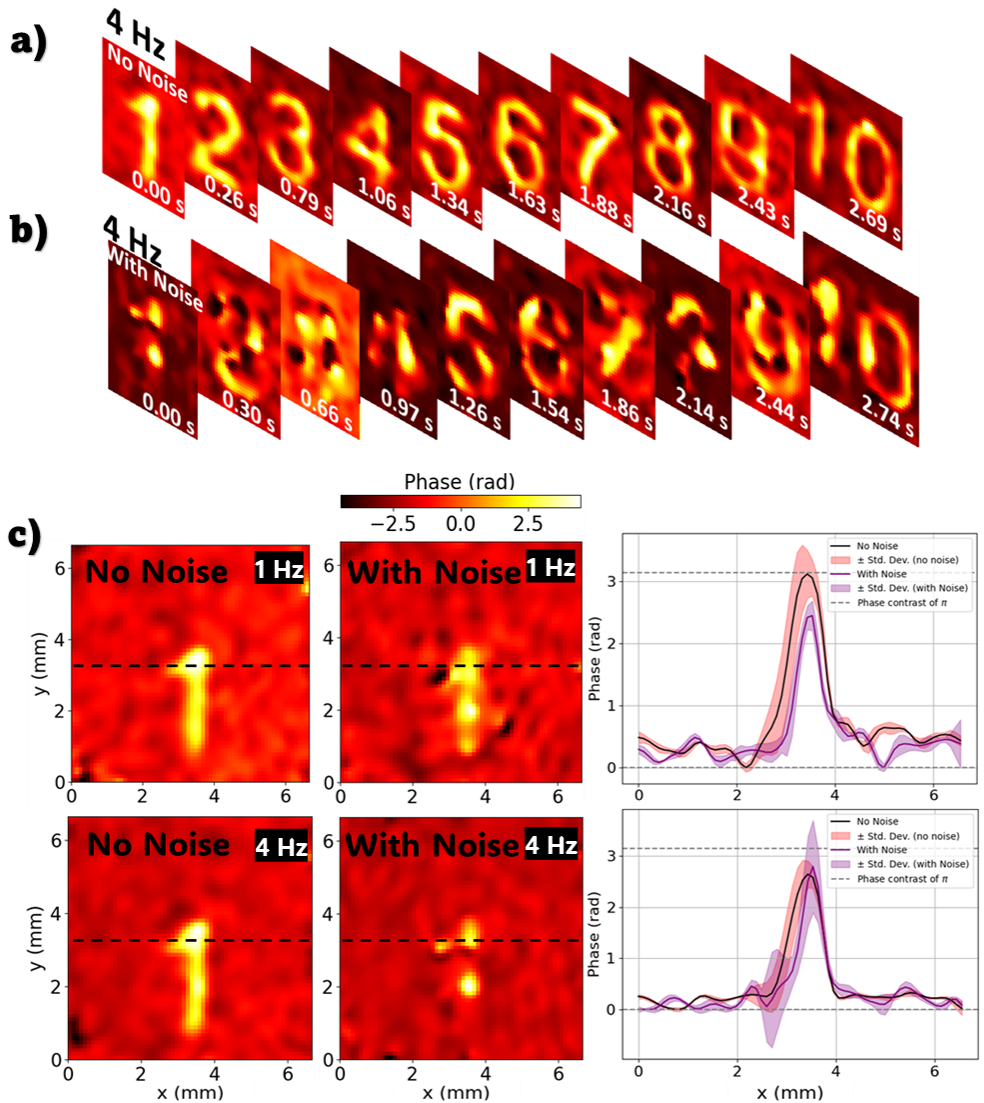}
\caption{Real-time phase imaging of sequential digits in the presence of noise. Sequences of digits from one to ten were projected onto the SLM to evaluate the performance of our scheme for real-time imaging under noisy conditions. (a) Reconstructed phase images of each digit and their corresponding time stamps at a projection rate of 4~Hz. (b) Phase reconstructions in the presence of noise, obtained with a signal-to-noise ratio of 1:5. (c) Comparison of the phase contrast for the digit one under different projection rates and in the presence or absence of noise.}
\label{fig:rates}
\end{figure}

The first sequence of SLM masks consists of 8-bit images in which the digits from one to ten, each displayed with a gray value of 128 on a black background corresponding to a gray value of 0. This sequence represents different object features, each encoded with a fixed phase value of 3.14~rad ($\pi$~rad). 
Figures~(\ref{fig:rates}a) and~(\ref{fig:rates}b) show the time stamps corresponding to the reconstructed phase patterns projected by the SLM at 4~Hz, without and with noise, respectively. The experiment demonstrates that our scheme successfully performs real-time phase-imaging distillation even under noisy conditions.

Figure~(\ref{fig:rates}c) shows a comparison of the phase reconstructions of the digit one at 1~Hz and 4~Hz. The line-profile plot compares the phase contrast between the phase images reconstructed with noise (purple) and without noise (black). The line profile is extracted from the dashed black line crossing the digit one. The standard deviation is calculated by considering the three rows above and below the dashed black line and is shown as the red (no noise) and purple (with noise) shaded regions surrounding the solid lines. In all cases, the digit one remains clearly distinguishable; however, the measured phase contrast shows a degradation due to both the projection rate and the presence of noise, deviating from the expected value of 3.14~rad.

\subsubsection{Dynamic phase contrast over time}

\begin{figure}[ht!]
\centering\includegraphics[width=0.9\textwidth]{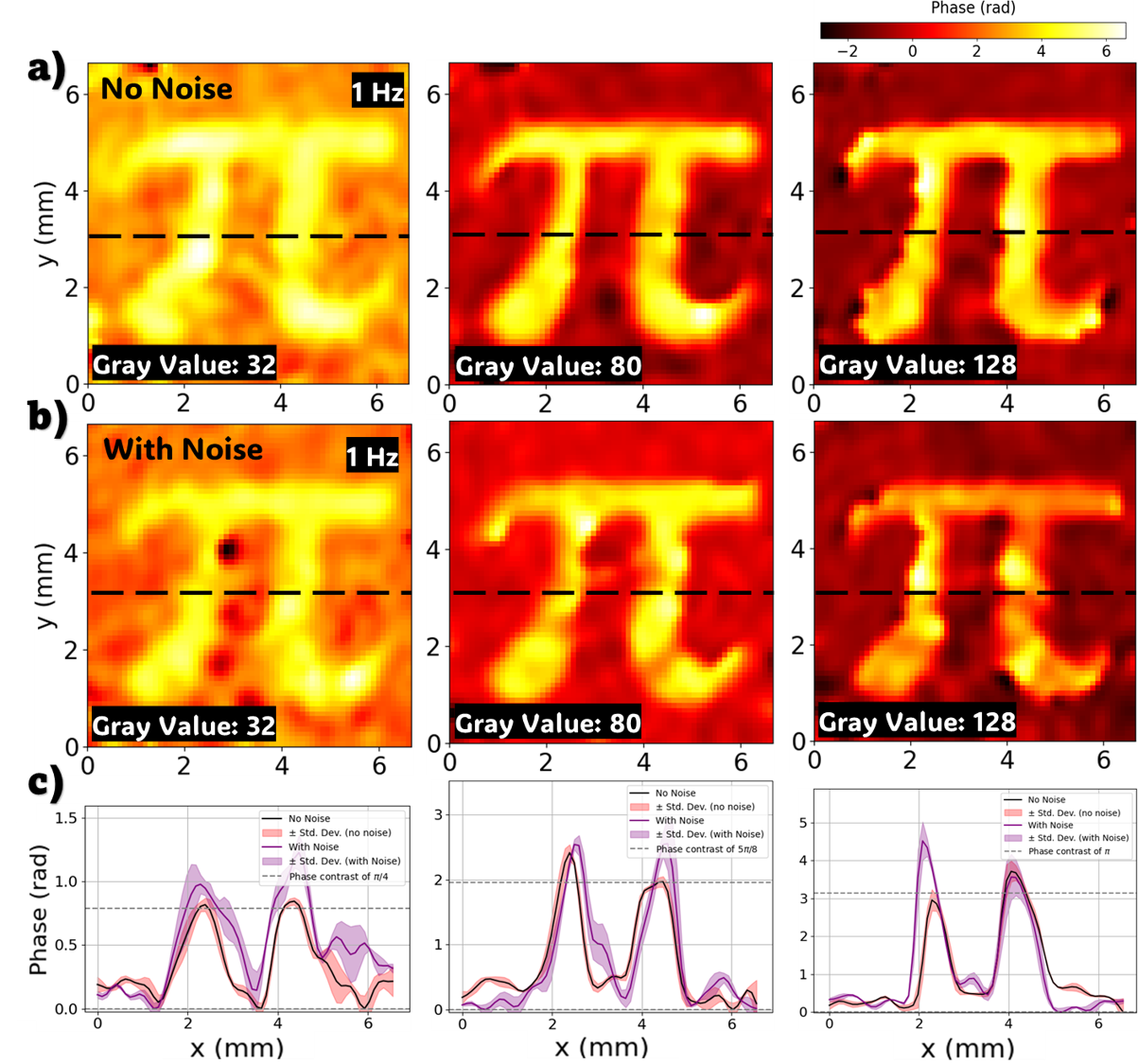}
\caption{Dynamic phase imaging of the Greek letter $\pi$ at 1~Hz with and without noise. (a) Three representative phase images corresponding to gray values of 32, 80, and 128, acquired without noise. (b) The same images acquired in the presence of noise. (c) Line-profile plots extracted from the dashed black line, comparing the phase contrast for each gray value with and without noise.}
\label{fig:Pi1}
\end{figure}

The second sequence of SLM masks consists of eight images of the Greek letter $\pi$, displayed with gray values varying from 16 to 128 in equal steps. This sequence simulates the dynamic evolution of the phase over time, going from $\frac{\pi}{8}~\text{rad}$ to $\pi$~rad (see the corresponding supplementary videos). Three representative images were selected to demonstrate the performance of our technique in imaging a time-evolving phase under noisy conditions. The reconstructed phase images correspond to gray (phase) values of 32 $\left(\frac{\pi}{4}~\text{rad}\right)$, 80 $\left(\frac{5\pi}{8}~\text{rad}\right)$, and 128 ($\pi$~rad).
Figure~(\ref{fig:Pi1}) compares the performance of our technique at a projection rate of 1~Hz, with and without noise.

Figures~(\ref{fig:Pi1}a) and~(\ref{fig:Pi1}b) show the evolution of the reconstructed phase images, transitioning from $\frac{\pi}{4}$ to $\pi$ over a time span of approximately 6~s (see the supplementary video for the full eight-image sequence). In Fig.~(\ref{fig:Pi1}b), the phase images are reconstructed in the presence of noise. The dashed black line indicates the region from which the line profiles shown in Fig.~(\ref{fig:Pi1}c) are extracted. Our technique successfully retrieves the features and contrast of the phase object as it dynamically evolves over time, with deviations in phase contrast observed due to the presence of noise.

\begin{figure}[ht!]
\centering\includegraphics[width=0.9\textwidth]{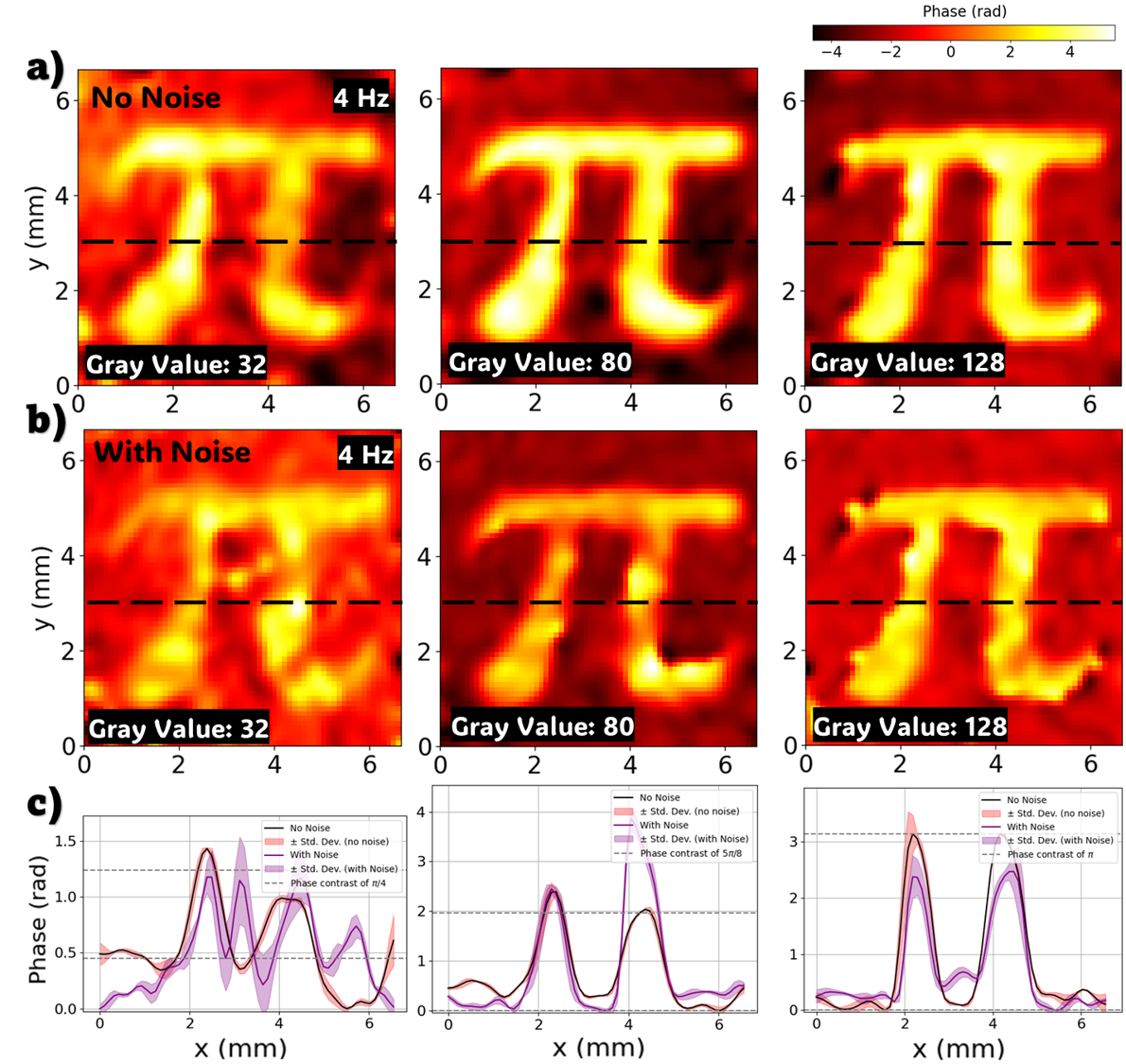}
\caption{Dynamic phase imaging of the Greek letter $\pi$ at 4~Hz with and without noise. (a) Three representative phase images corresponding to gray values of 32, 80, and 128, acquired without noise. (b) The same images acquired in the presence of noise. (c) Line-profile plots extracted from the dashed black line, comparing the phase contrast for each gray value with and without noise.}
\label{fig:Pi4}
\end{figure}

Similarly, Fig.~(\ref{fig:Pi4}) compares the performance of our technique at a projection rate of 4~Hz, with and without noise. Figures~(\ref{fig:Pi4}a) and~(\ref{fig:Pi4}b) show the evolution of the reconstructed phase images, transitioning from $\frac{\pi}{4}$ to $\pi$ over a time span of approximately 1.5~s (see the supplementary video for the full eight-image sequence). Even at this higher projection rate, the system successfully retrieves a phase contrast close to the expected value (see the solid black line and dashed gray line in Fig.~(\ref{fig:Pi4}c)). When noise is introduced, larger deviations in the measured phase contrast are observed. These deviations are attributed to the combined effects of noise and the higher projection rate.
 
\begin{figure}[ht!]
\centering\includegraphics[width=0.45\textwidth]{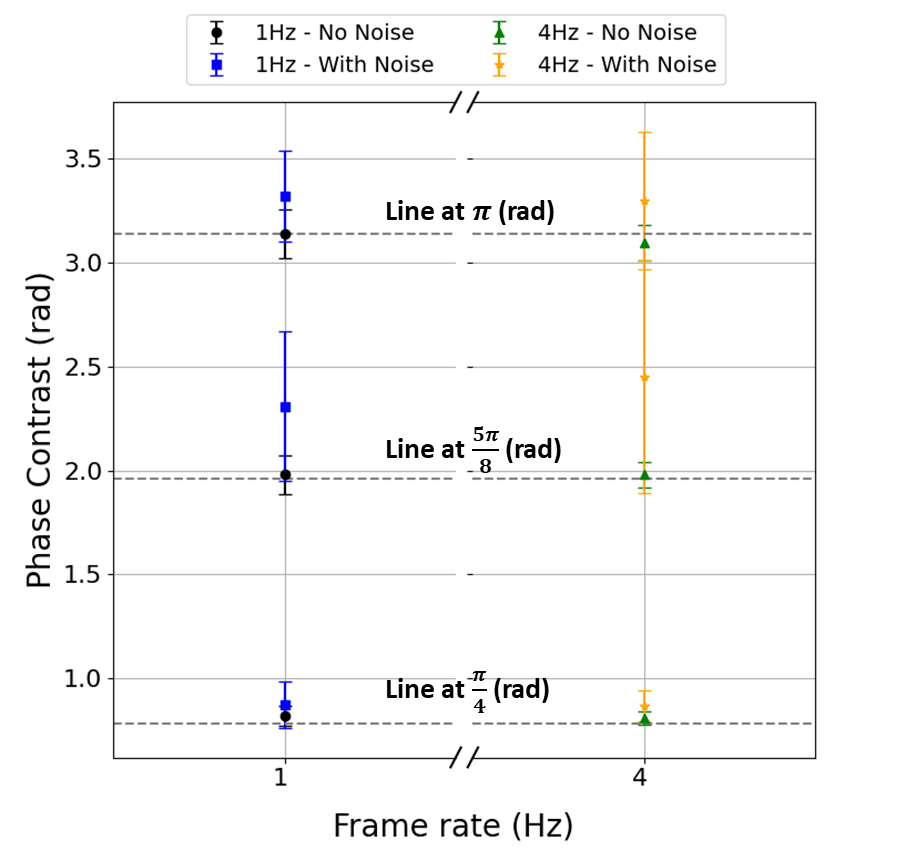}
\caption{Phase contrast for different gray values of the SLM masks measured at 1~Hz and 4~Hz, corresponding to Figs.~(\ref{fig:Pi1}) and~(\ref{fig:Pi4}), respectively. The dashed-lines indicate the expected phase values for the three displayed cases: $\frac{\pi}{4}$ rad for 32 gray-value, $\frac{5\pi}{8}$ rad for 80 gray-value and $\pi$ rad for 128 gray-value.}
\label{fig:Quant2}
\end{figure}

In addition, Fig.~(\ref{fig:Quant2}) illustrates the performance of our technique by showing the measured phase contrast for different gray values at both frame rates. As expected, the results show that the largest errors occur under noisy conditions compared to noise-free operation, and that the frame rate of the SLM masks also influences the measurement accuracy. In particular, a frame rate of 4~Hz introduces errors larger than 1~Hz.
The phase contrast values shown in Fig.~(\ref{fig:Quant2}) are extracted in a similar manner to Fig.~(\ref{fig:Quant1}), using three evenly spaced line profiles across the reconstructed phase images and extracting four phase values from each of them. 

To conclude, we compare in Table (1) our noise-resilient method with all currently existing techniques that rely on photon pairs to perform image distillation, in other words imaging under noisy conditions. While previous approaches, such as quantum illumination protocols \cite{gregory_imaging_2020, gregory_noise_2021}, coincidence-based schemes \cite{defienne_quantum_2019}, and QIUL-based image distillation \cite{fuenzalida2023}, have demonstrated impressive noise rejection capabilities, they typically require long integration times, multi-frame acquisition, or statistical post-processing. In contrast, our method enables single-shot, real-time amplitude and phase imaging using undetected light, and remains functional even when the noise intensity exceeds the signal by more than an order of magnitude. This positions our technique as a significant advance toward practical, high-speed quantum imaging in realistic environments.

\begin{table}[ht!]
\centering
\footnotesize

\renewcommand{\arraystretch}{1.4}
\begin{tabular}{p{2.0cm} p{2.65cm} p{1.0cm} p{2.5cm}}
\hline
\textbf{Publication} & \textbf{Methodology} & \textbf{SNR} & \textbf{Acquisition Time} \\
\hline

\textit{Sci. Adv.} (2019) 
& Photon-correlations and coincidence detection & 1:10 & $10^{7}$ frames of 6~ms $\approx$ 16.6~h \\

\textit{Sci. Adv.} (2020) 
& Photon-correlations and coincidence detection  & 1:5.8 & $2.5*10^{6}$ frames of 15~ms $\approx$ 10.4~h \\

\textit{Sci. Rep.} (2021) 
& Photon-correlations and coincidence detection & 1:20 & $2.4*10^{6}$ frames of 14.95~ms $\approx$ 10~h \\

\textit{Sci. Adv.} (2023)
& QIUL + DPSH & 1:252 & 12 frames of 1~s = 12~s \\

\textbf{This work} 
& \textbf{QIUL + OAH} & \textbf{1:12} & \textbf{1 frame of 0.25~s = 0.25~s} \\

\hline
\end{tabular}
\caption{Comparison of quantum imaging techniques in noisy environments.}
\label{tab:comparison}
\end{table}

\newpage

\section{Discussion}

We have introduced a technique for performing real-time imaging in noisy environments via undetected light, enabling the reconstruction of both amplitude and phase information using photons that never interact with the sample. Here, real-time imaging denotes the capability of the method to follow a dynamic sample as its spatial features change, with a temporal resolution sufficiently high to capture these variations as they occur. The approach allows us to achieve frame rates of up to 4~Hz. Although slower than video-rate imaging (30~Hz), these frame rates are sufficient to track the dynamic phase patterns used in our experiment and represent a significant advance compared to multi-second or multi-minute integration times typical of existing quantum imaging techniques. We have separated the spatial information carried by quantum and classical light sources by employing QOAHUL \cite{Leon-Torres(2024), Pearce2024}, as illustrated in Fig.~(\ref{fig:Distillation}). The core mechanism responsible for noise resilience arises from quantum Fourier off-axis holography \cite{Topfer:25, Leon-Torres(2024), Pearce2024}, implemented within a hybrid nonlinear interferometer \cite{Leon-Torres(2024), Kim2024} based on induced coherence without induced emission \cite{Mandel1991, Grayson1994}. 

This approach opens new possibilities for implementing quantum imaging techniques in real-world scenarios where the presence of classical noise would otherwise make image reconstruction unfeasible. Furthermore, the ability to perform real-time imaging enhances its applicability to samples whose phase varies rapidly over time. Our technique is capable of retrieving the object information as long as the reconstructed phase contrast remains within the expected phase value, even when the noise intensity is up to twelve times higher than the signal intensity. Beyond this level, increasing noise progressively degrades the fringe contrast and ultimately renders the reconstruction unfeasible.  Figure~(\ref{fig:Quant1}) demonstrates the ability of the system to extract phase contrast information across different noise levels. 

In our experiment, we employed a static classical noise source, whose spatial frequencies in Fourier space remain constant. Our distillation technique is also applicable to spatially structured noise sources with dynamic Fourier components. This condition holds as long as the spatial-frequency components corresponding to the object information remain sufficiently dominant relative to the noise contributions and can be spectrally filtered. Such filtering is achieved by controlling the relative angle between the object and reference beams, introducing a linear phase through mirror tilting. 

The main limitation of our approach arises from the contrast of the interference fringes. The spatial overlap of the hologram and the noise at the camera plane leads to a degradation of the fringe contrast; the larger the noise, the lower the resulting interference contrast. Under these conditions, our technique becomes unable to distinguish between the noise and the object spatial frequencies, preventing effective spectral filtering and ultimately limiting the signal-to-noise ratio range for which the method remains applicable.

Furthermore, we investigated the capabilities of our setup for dynamic phase imaging in the presence of noise. These results are presented in Figs.~(\ref{fig:rates})–(\ref{fig:Pi4}), where two image sequences were acquired at frame rates of 1 and 4~Hz, with a signal to noise ratio of 1:5. Our setup reproduces the phase contrast with good accuracy, even as the phase changes rapidly over time, as shown in Fig.~(\ref{fig:Quant2}). While the presence of noise and higher projection rates lead to larger deviations, the main object features are nonetheless preserved. We attribute these deviations to a temporal lag between the SLM projection rate and the image acquisition process, which slightly distorts the reconstructed phase during fast temporal modulations.


The promising results obtained in our experiments point the way toward further improvements, particularly in achieving higher projection rates while maintaining accurate phase measurements. This limitation arises primarily from imperfect synchronization between the camera exposure time and the SLM projection rate. Such mismatch can lead to averaging of frames that do not correspond to the same phase pattern, resulting in image distortions and deviations in the reconstructed phase. To mitigate this, several improvements can be implemented. A faster data-processing pipeline would reduce latency between acquired frames, while optimizing the initial image size and the recursive spectral-filtering functions used to isolate the spatial frequencies of the object in Fourier space would further speed up reconstruction and enable shorter exposure times. Beyond algorithmic optimization, the hologram reconstruction stage, currently a major bottleneck for timely data analysis, could be substantially accelerated by integrating deep neural networks \cite{Liu2025}. Another limiting factor is the optical pump power of the nonlinear crystal (60~mW in our case), which constrains the minimum achievable exposure time and therefore the maximum projection rate of the SLM when simulating dynamic phase behavior. We believe that once these technical refinements are implemented, our approach could serve as a bridge toward the deployment of quantum imaging schemes in real-world scenarios ranging from quantitative phase imaging to manufacturing.

\begin{backmatter}
\bmsection{Funding}

The authors acknowledge support by the Carl-Zeiss-Stiftung within the Carl-Zeiss-Stiftung Center for Quantum Photonics (CZS QPhoton) under the project ID P2021-00-019, and the Deutsche Forschungsgemeinschaft (DFG, German Research Foundation) under Germany´s Excellence Strategy – EXC 2051 – Project-ID 390713860. 

In addition, this work was supported by the Horizon WIDERA 2021-ACCESS-03-01 grant 101079355 "BioQantSense", from the European Union’s Horizon 2020 Research and Innovation Action under Grant Agreement No. 101113901 (Qu-Test, HORIZON-CL4-2022-QUANTUM-05-SGA) and by grants funded by the Federal Ministry of Research, Technology and Space (QUANTIFISENS, 03RU1U071M and QUANCER, 13N16441).

\bmsection{Acknowledgments}

We would like to show our gratitude to V. Kaipalath and C. Sevilla for sharing their time and knowledge with J.R.L.T. during the course of this research.

\bmsection{Disclosures}

The authors declare no conflict of interest.

\bmsection{Data availability}

The data that support the findings of this study are available from the corresponding author, J.R.L.T., upon reasonable request.

\bmsection{Supplemental document} 

See Supplementary videos for supporting content.

\end{backmatter}

\bibliography{Optica-template}

\end{document}